# Elastic energy and string configurations in the chiral gauge theory of biaxial-uniaxial nematic phase transitions

L. V. Elnikova

Institute for Theoretical and Experimental Physics, Moscow 117218, Russian Federation

**Abstract.** In nematic liquid crystals (NLCs), topological defects of a chiral origin play a role in phase transitions and lead to phase configurations of nontrivial topology, like those in neutron stars and helium in the A-phase. In the biaxial-uniaxial phase transition, the deformation of the orbit from $SO(3)/(Z_2 \times Z_2)$ to $SO(3)/Z_2$, as the order parameter degeneracy of the NLC, connects together an evolution of topological defects, the surface anchoring energy and elastic Frank modui. In this work we estimate the chiral gauge field presentation of the constrained Ladnau-de Gennes theory of the biaxial nematics, which have to explain their topologically dependent phase transformation, using the description of the transformation of disclinations in the biaxial nematic phase into the surface bojooms of a uniaxial NLC.

**Introduction**

Nematic liquid crystals (NLCs) have wide applications in modern spintronic, micro-fluidic and optic devices, sensors and actuators *etc*. [1,2] thanks to their ability to realign molecules in the phase transitions. Therefore, studies of NLC aggregations for that goals and predictions of their physical properties are motivated.

NLC are characterized by an orientational ordering of molecules described by the vector director. The uniaxial nematic $N_u$ LCs have a single director ***n***=−***n*** coinciding with the symmetry axis and the optic axis of a crystal. The biaxial nematic $N_b$ phase is characterized by three mutually orthogonal vectors, i.e. by one primary ***n***=−***n*** and two secondary ***m***=−***m***, ***l***=−***l*** directors.

Biaxial nematics have been discovered in lyotropic LC (potassium laurate/1-decanol/water system) and are often observable in colloid, emulsive and micellar LC systems generally with optical spectroscopy (fluorescence confocal polarizing microscopy and coherent anti-Stokes Raman scattering *etc*). In thermotropic LCs, the $N_b$ phase may appear in pentylcyanobiphenyl (5CB), C7, C12, A131, in organo-siloxane tetrapodes, in the bent-core liquid crystal thiadiazole (DT6Py6E6) [3]) *etc*. So, reversible biaxial to uniaxial nematic phase transitions $N_b \leftrightarrow N_u$ were observed recently in DT6Py6E6 [1,3] and in E7 with light microscopy (3D), via observations of the isogyre's splitting.

The elastic constants together with the anchoring energy at the interface may be determined in a torsion pendulum experiment in magnetic field ($K_{ii}$ in 5CB [4] and references therein) by measuring the threshold value of the magnetic field for the Freedericks transition in thin nematic layers.

The macroscopic theory described the sequence of $N_b \to N_u \to I$(isotropic) transitions and topological defects in NLCs is well developed [5-11] in spirit of the Landau-de-Gennes (LdG) and the Oseen-Frank theories. The elastic moduli of the LdG theory and coefficients of viscosity are attributed to intermolecular interactions of NLCs and may be comprehended with relaxation characteristics [12]. The free energy density of NLCs depends on three "bulk" Frank elastic constants, i.e., splay $K_{11}$, twist $K_{22}$ and bend $K_{33}$ (of the order of $10^{-6}$-$10^{-7}$ dynes), and two "surface" mixed elastic constants splay-bend $K_{13}$ and saddle-splay $K_{24}$, the last constant is addressed to the surface anchoring.

Different types of point and linear topological defects which might arise, evolve, interact, annihilate or disappear in NLC phases is discussed in [2,4] *etc*. Such phase transformations are often formulated in frames of the Kibble-Zurek theory, and their explanations may be described as singularities like monopoles, strings, D-branes, domain walls *etc*. The appearance of topological defects is defined by symmetry of a NLC, by surface anchoring and by the relative values of elastic constants [13].

In [14], the authors showed, that in droplet NLC's, point topological defects (hedgehogs) evolving form a radial hedgehog to a hyperbolic one at the phase transition accompanied by lowering of the symmetry $K \to C_{\infty,h}$, are associated with the temperature-induced changes in the elastic constants.

In this paper, we analyzed the gauge versions of the LdG and the Oseen-Frank theories to the $N_b \to N_u$ transition, when the order parameter coincides with the deformation of the orbit from $SO(3)/(Z_2 \times Z_2)$ to $SO(3)/Z_2$, this example allows us to explore the influence of the chiral symmetry breaking and the some topological defect's evolution onto the free energy and other thermodynamical parameters of NLCs in the $N_b \to N_u$ transition.

**Constrained Oseen-Frank theory and chiral gauge theory**

The Oseen-Frank theory of $N_b$'s and $N_u$'s operates with the order parameter, which is expressed in terms of magnetic susceptibility $Q$, this constrained biaxial symmetric traceless $Q$ tensor have distinct constant eigenvalues $\lambda_1$, $\lambda_2$, $\lambda_3$, it is of the non-abelian eight-element quaternion group [5,16].

As it mentioned in [5,6,7], in the $N_b \to N_u$ transition lowing the symmetry of a NLC, the group orbits reflect the order parameter space of a NLC, which are classified in the full homotopic sequence of the fundamental groups $\pi_i$ ($i=0, 1, 2$) [5]; so there the full energy of a NLC, especially, the anchoring term, depend on the group symmetry of a NLC, in which only certain defects are enable to stabilize or to transform. The deformation of the non-trivial orbit $SO(3)/(Z_2 \times Z_2)$ to $SO(3)/Z_2$, was wide studied for a long time analytically [5,6,7] and numerically ([13] and references therein), for the case of the semidirect product $Z_2 \times Z_2$, we have a dihedral group $D_{2h}$.

Let us consider volume disclinations in droplets of the $N_b$ nematic, which became bojooms of the $N_u$ phase in the $N_b \to N_u$ transition (Fig. 1).

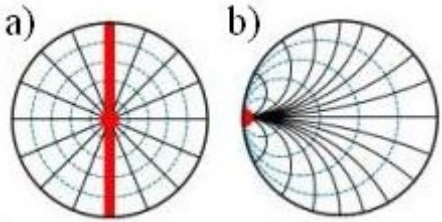

Fig. 1 The director's configurations at perpendicular anchoring (black lines denote the primary director, blue one is the secondary director) and a) the $N_b$ phase with a disclination (a red line), b) the phase with a bojoom (a red dot).

In the gauge field theory, the $SO(3)$ group is locally isomorphic to $SU(2)$ and may be presented as $SU(2)/Z_2$. Disclinations and bojooms of physical systems may be identical to strings and monopoles, respectively. When a non-abelian string disappears (if we have a linear singularity, the group $\pi_1(S^2)=Z_2$, a disclination of a $N_b$ is unstable), it may become a bojoom depending on its topological characteristics, the strength $m$, the charge $N$ and a class of coupled elements $C_i$ of the group $Q$ [11].

The free energy of a $N_b$ is expressed ([16] and Refs. therein) as follows:

$$F(\mathbf{Q}, \Omega) := \int_\Omega [\psi(\mathbf{Q}, \nabla \mathbf{Q}) + f_B \mathbf{Q})] dx, \tag{1}$$

where $f_B(\mathbf{Q}) = f_B(tr(\mathbf{Q}^2), tr(\mathbf{Q}^3))$ is the bulk free-energy density (a function of the principal invariants of $Q$) and $\psi(\mathbf{Q}, \nabla \mathbf{Q}) = L_1 I_1 + L_2 I_2 + L_3 I_3 + L_4 I_4$ is the elastic free-energy density, where the $L_i$ are material constants and $I_i$ are the elastic invariants $I_1 = \mathbf{Q}_{ij,k}$; $I_2 = \mathbf{Q}_{ik,j}\mathbf{Q}_{ij,k}$; $I_3 = \mathbf{Q}_{ij,k}\mathbf{Q}_{ij,k}$; $I_4 = \mathbf{Q}_{lk}\mathbf{Q}_{ij,l}\mathbf{Q}_{ij,k}$. $f_B(Q)$ is invariant under the $SO(3)$-action by conjugation on the five-dimensional space of $Q$-tensors, so that the critical points of the bulk energy form an orbit of solutions in the five-dimensional space of $Q$-tensors [16]. $Q$ is assumed to have constant scalar order parameters $S_1$ and $S_2$, and hence constant eigenvalues $\lambda_1=(2S_1-S_2)/3$, $\lambda_2=(2S_2-S_1)/3$, $\lambda_1=(S_1+S_2)/3$; ($\lambda_1 \leq \lambda_2 \leq \lambda_3$). In the constrained theory, the bulk energy is constant, and we calculate only the elastic free energy.

A few presentations of the functional corresponding to the energy density given by the elastic invariant [16] allow us to compute the biaxial NLC's directors $n$, $m$, $l$ ($|\nabla n|^2, |\nabla m|^2, |\nabla l|^2$) in terms of the derivatives of the vector map $\Omega \to S^3$. We may limited by the third elastic invariant $I_3 = Q_{ij,k}Q_{ij}$, and deduce the free energy functional into the expression:

$$\mathbf{Q} \bullet \quad I_3(\mathbf{Q}) = \int_\Omega I_3(\mathbf{Q}, \nabla \mathbf{Q}) dx, \tag{2}$$

which is well defined and finite on the Sobolev class $W^{12}(\Omega, Q(\lambda_1, \lambda_2, \lambda_3))$, summing to our consideration the Sobolev class $W^{1p}(\Omega, Q(\lambda_1, \lambda_2, \lambda_3)$ and the elastic free energy functional on $I_3$.

Constructing a simple general scenario of a decay of a line disclination into a bojoom (the surface point defect) of a $N_b$ droplet at the $N_b \to N_u$ transition, we may build the association: the open loop, thanks to an increasing of its tension transforms into a monopole.

On the other hand, we apply the follows calculation method. Due to the lattice gauge formulation and in the suitable differential forms on a dual lattice [17], we use the action from the form (3) in the link (site) variables $\theta(\varphi)$ for the introducing the operator # of the mutual gauge transformations between the plaquette and the link variables, # corresponds to the correct linking number of the currents of the lattice.

So we may calculate the second-rank order parameter $Q$ in the new charge variables ($\kappa, \beta, \mu$) on a simple cubic lattice, leaving as an illustration, the self-dual conditions (Fig. 2).

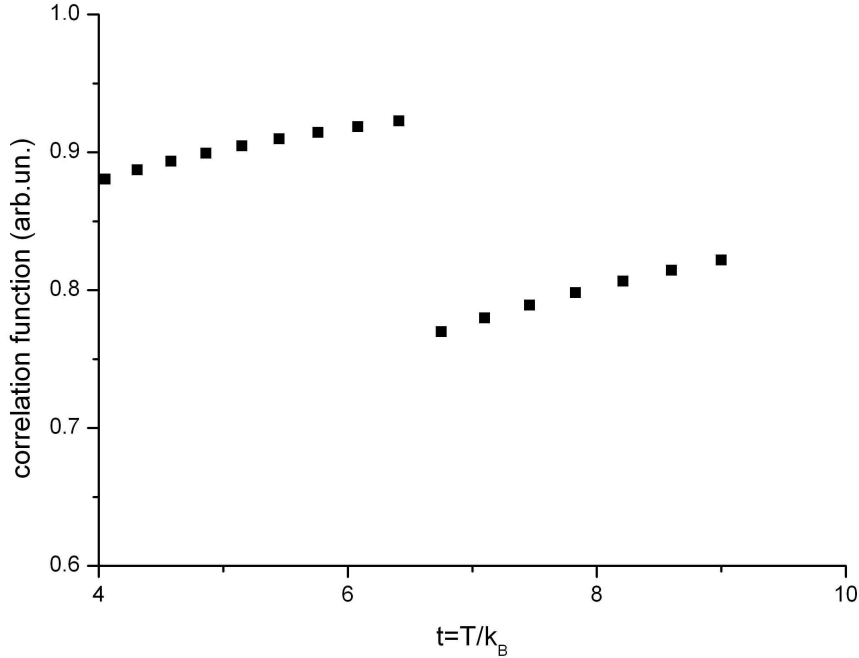

Fig. 2 The order parameter (arbitrary units) vs. temperature in the Monte Carlo tests (accuracy 5%) reflects the decay of a string on a monopole (a disclination of the $N_b$ phase to a bojoom of the $N_u$ phase at the $N_b \to N_u$ transition), showing the gap.

The partition function reads [18]

$$Z = const \sum_{\substack{*j \in Z(C_{k+2}) \\ \delta j = 0}} \exp\{-4\pi^2 \beta(j, \Delta^{-1} j)\}, \tag{3}$$

where the summation is going over the $(D-k-2)$ forms $j$ having the physical meaning of the "monopole current", and $C_k$ is the $k$-dimensional cell on the lattice. The gap in the order parameter (Fig. 2) gives an evidence on a "string" is broken and a monopole (monopoles) appeared.

Two elastic constants *K*, $K_{24}$ will be extracted due to (1).

**Summary**

In the observables of NLC's, it is important to have an explanation of the transformations of possible phases possessing of non-abelian topological defects into abelian ones and the role of elasticity in such transformations. We clarified, that the lattice gauge calculations is a helpful instrument to analyze the "string-monopole" configurations, corresponded to the characteristic defect transformations at the $N_b \rightarrow N_u$ transition in NLCs'. As these factors define physical properties of novel synthesizable NLC compounds, therefore similar estimations and predictions are required.

**References**

[1] Y.-K. Kim, B. Senyuk, O.D. Lavrentovich, Molecular reorientation of a nematic liquid crystal by thermal expansion, Nature communications 3 (2012) 1133-1139.

[2] Y.-K. Kim, B. Senyuk, S.-T. Shin, A. Kohlmeier, G.H. Mehl, O.D. Lavrentovich, Surface alignment, anchoring transitions, optical properties, and topological defects in the thermotropic nematic phase of an organo-siloxane tetrapodes, Soft Matter 10 (2014) 500-509.

[3] Y.-K. Kim, M. Majumdar, B.I. Senyuk, L. Tortora, J. Seltmann, M. Lehmann, A. Jákli, J.T. Gleeson, O.D. Lavrentovich, S. Sprunt, Search for biaxiality in a shape-persistent bent-core nematic liquid crystal, Soft Matter 8 (2012) 8880-8890.

[4] S. Faetti, M. Gatti, V. Palleschi, Measurements of surface elastic torques in liquid crystals: a method to measure elastic constants and anchoring energies, Revue Phys. Appl. 21 (1986) 451-461.

[5] M.I. Monastyrsky, Topology of gauge fields and condensed matter, Plenum, 1993.

[6] F. Gay-Balmaz, M. Monastyrsky, T.S. Ratiu, Lagrangian reductions and integrable systems in condenced matter, arXiv:1404.7654v1.

[7] M.I. Monastyrsky, P.V. Sasorov, Topology of the lattice of vortices in neutron stars, Modern Physics Letters A 26(4) (2011) 267-277.

[8] E. Govers, G. Vertogen, Elastic continuum theory of biaxial nematics, Phys. Rev. A. 30 (1984) 998-2000.

[9] J.P. Straley, Ordered phases of a liquid of biaxial particles, Phys. Rev. A. 10 (1974) 1881-1887.

[10] N.D. Mermin, The topological theory of defects in ordered media, Reviews of Modern Physics, 51 (1979) 591-648.

[11] M.V. Kurik, O.D. Lavrentovich, Defects in liquid crystals: homotopy theory and experimental studies, Sov. Phys. Usp. 31 (1988) 196-224.

[12] G. Barbero, E.K. Lenzi, Importance of the surface viscosity on the relaxation of an imposed deformation in a nematic liquid crystal cell, Physics Letters A 374 (2010) 1565-1569.

[13] C. Chiccoli, I. Feruli, S.V. Shiyanovskii, O.D. Lavrentovich, P. Pasini, C. Zannoni, Topological Defects in Schlieren Textures of Biaxial and Uniaxial Nematics, Phys. Rev. E. 66 (2002) 030701-1-4.

[14] O.D. Lavrentovich, E.M. Terent'ev, Phase transition altering the symmetry of topological point defects (hedgehogs) in a nematic liquid crystal, *Sov.* Zhurnal Eksperimentalnoi I Teoreticheskoi Fiziki 91(6) (1986) 2084-2096.

[16] D. Mucci, L. Nicolodi, On the elastic energy density of constrained Q-tensor models for biaxial nematics, Arch. Ration. Mech. Anal. 2006 (2012) 853-884.

[17] T.L. Ivanenko, M.I. Polikarpov, Symmetries of the Chern-Simons theory on the lattice, Nucl. Phys. B (Proc. Suppl) 26 (1992) 536-538.

[18] T.L. Ivanenko, M.I. Polikarpov, Symmetries of Abelian lattice theories with Chern-Simon interactions, Preprint ITEP-49, Moscow, 1991.